\documentclass[a4paper,fleqn,usenatbib]{mnras}

\usepackage{newtxtext,newtxmath}
\usepackage[T1]{fontenc}
\usepackage{ae,aecompl}
\usepackage{graphicx}	% Including figure files
\usepackage{amsmath}	% Advanced maths commands
\usepackage{breqn}
\usepackage{hyperref}
\newcommand{\be}{\begin{eqnarray}}
\newcommand{\ee}{\end{eqnarray}}

\title[Constraints on EMDA gravity from the iron line]{Constraints on Einstein-Maxwell dilaton-axion gravity\\from X-ray reflection spectroscopy}

\author[Tripathi et al.]{Ashutosh~Tripathi,$^{1}$
Biao~Zhou,$^{1}$
Askar~B.~Abdikamalov,$^{1,2}$ \newauthor
Dimitry~Ayzenberg,$^{3}$
and Cosimo~Bambi$^{1}$\thanks{Corresponding author: \href{mailto:bambi@fudan.edu.cn}{bambi@fudan.edu.cn}} \\
$^{1}$Center for Field Theory and Particle Physics and Department of Physics, Fudan University, 200438 Shanghai, China\\
$^{2}$Ulugh Beg Astronomical Institute, Tashkent 100052, Uzbekistan\\
$^{3}$Theoretical Astrophysics, Eberhard-Karls Universit\"at T\"ubingen, 72076 T\"ubingen, Germany
}

\begin{document}
\label{firstpage}
\pagerange{\pageref{firstpage}--\pageref{lastpage}}
\maketitle

\begin{abstract}
Einstein-Maxwell dilaton-axion gravity is a string-inspired model arising from the low energy effective action of heterotic string theory and an important candidate as alternative to General Relativity. Recently, some authors have explored its astrophysical implications in the spectra of accreting black holes and inferred the constraint $r_2 < 0.1$, where $r_2 \ge 0$ is the black hole dilaton charge and General Relativity is recovered for $r_2 = 0$. In the present paper, we study the impact of a non-vanishing black hole dilaton charge on the reflection spectrum of the disk. From the analysis of a \textsl{NuSTAR} spectrum of the black hole binary EXO~1846--031, we find the constraint $r_2 < 0.011$ (90\% CL), which is an order of magnitude more stringent.
\end{abstract}


%%%%%%%%%%%%%%%%%%%%%%%%%%%%%%%%%%%%%%%%%%%%%%%%%%

%%%%%%%%%%%%%%%%% BODY OF PAPER %%%%%%%%%%%%%%%%%%

\section{Introduction}

General Relativity (GR) was proposed by Albert Einstein in 1915 and has become one 
of the most profound theories in physics. It has been applied to various physical phenomena and has been tested in numerous ways in the weak field regime \citep{2014LRR....17....4W}. Over the past few years, tests in the strong field regime of gravity have begun \citep{book, 2017RvMP...89b5001B, 2019PhRvD.100j4036A} thanks to advancements in instrumentation and technology. However, GR has been found to have limitations both in theoretical (e.g., singularities, hierarchy problem) and observational (e.g., dark energy, dark matter) aspects which motivate the search for a modified theory of gravity that could resolve these issues.

Various modified theories have been proposed in order to address these limitations. Many of these theories are motivated by attempts to quantize gravity, e.g.~string theory \citep{2020arXiv200401210C}. In order to test GR, it is necessary to explore whether a quantized theory could be tested with low energy observations, e.g.~Einstein-Maxwell dilaton-axion (EMDA) gravity arises from the low energy Lagrangian of superstring theories \citep{1992PhRvL..69.1006S, 2002CQGra..19.5063R}. In this case, the ten dimensional string theory is compactified on a six dimensional torus. This theory includes pseudo-scalar axion and scalar dilaton fields coupled with the Maxwell field and metric, leading to significant observational implications \citep{2006PhRvD..74j3508S, 2008JCAP...03..012C}. Thus, this theory could be used to study the spacetime metric with astrophysical observations as done in the past \citep{2007PhRvD..75b3006G, 2018NatAs...2..585M, 2018PhRvD..97b4003A, 2016PhRvD..94h4025Y, 2008PhRvD..78d4007H, 2020arXiv200212786N}.

Black holes (BHs) are the most extreme gravitational objects known of in the Universe. In GR, all isolated, stationary, and axisymmetric astrophysical (uncharged) BHs are expected to be described by the Kerr solution, in which case they are fully described by only the mass $M$ and spin parameter $a_*$ \citep{1975PhRvL..34..905R, 1967PhRv..164.1776I, 1968CMaPh...8..245I, 1972CMaPh..25..152H, 1971PhRvL..26..331C}. Since BHs in GR are expected to be such simple objects, departures from the Kerr solution would, in principle, have clear signatures in observations of BHs. Moreover, the population of BHs in our Galaxy and the Universe is expected to be quite significant. Most, if not all, galaxies are expected to harbor a supermassive BH at their center and several dozen stellar-mass BHs have been found through electromagnetic observations with even more observed through gravitational wave detections \citep[see, e.g.,][]{2020mbhe.confE..28B}. The simplicity of BHs in GR and their not insignificant population make them great laboratories for testing gravity.

Gravitational waves, X-ray reflection spectroscopy, and BH shadow observations are some of the leading techniques which use BHs as probes for testing GR. In this work, we will focus on X-ray reflection spectroscopy, the study of the relativistic reflection spectrum from the inner part of the accretion disk around BHs to study the properties of the spacetime \citep{2014SSRv..183..277R,2020arXiv201104792B}. Material from the host galaxy (for supermassive BHs) or from a binary companion (for X-ray binaries) is gravitationally pulled into orbit around the BH and forms an accretion disk. The disk is generally modeled as infinitesimally thin, optically thick, and made up of differentially rotating layers that, through viscous torques between the layers, transfer angular momentum outwards and heat inwards. The hot disk radiates thermally, and some of this radiation can be inverse Compton scattered by a nearby corona, i.e.~a hotter optically thin cloud of gas. Some of the up-scattered radiation can return to the disk and be reflected, producing what is known as the reflection spectrum. The reflected radiation of a number of BHs has been detected by X-ray telescopes and analyzed to study the properties of BHs and emission from the accretion disk.

The inner region of the accretion disk is an ideal astrophysical laboratory to test GR with observations. The relativistic reflection coming from this particular region is strongly dependent on the BH spacetime and can be used as a promising feature to study possible deviations from GR. In EMDA gravity, the Kerr-Sen metric describes a rotating and axi-symmetric BH, analogous to the Kerr BH in GR. The current state-of-the-art relativistic reflection model \textsc{relxill} \citep{2014ApJ...782...76G, 2014MNRAS.444L.100D} describes reflection from the accretion disk assuming the Kerr BH solution within GR. It has been extended to other spacetime metrics in the \textsc{relxill\_nk} suite of models \citep{2017ApJ...842...76B,   
2018PhRvL.120e1101C, 2018PhRvD..98b3018T,  2019ApJ...878...91A, 2019EL....12530002Z,  2020PhRvD.101f4030T, 2020PhRvD.102j3009T, 2020EPJC...80..400Z}. These \textsc{relxill\_nk} modified gravity theory models are used to constrain the deviations from the Kerr solution using X-ray observations of BHs \citep{2020PhRvD.102l4071N, 2019PhRvD..99l3007L, 2018ApJ...865..134X, 2019ApJ...879...80C, 2018PhRvD..98b4007Z, 2019ApJ...874..135T, 2019ApJ...875...56T, 2019ApJ...875...41Z, 2019ApJ...884..147Z, 2019PhRvD..99h3001T, 2020arXiv201210669T}.

In this work, we have developed a relativistic reflection model within the \textsc{relxill\_nk} framework using the BH solutions given by EMDA gravity. We have analyzed the 2019 \textsl{NuSTAR} observation of an X-ray binary EXO~1846--031 studied in~\citet{2020ApJ...900...78D} and placed constraints on the deformation parameter present in the theory. We have also employed MCMC simulations to study the degeneracy between the various physical parameters and computed the error on the parameters.

The content of this paper is as follows. 
In Section~\ref{s-emda}, we describe the Kerr-Sen BH solution in EMDA gravity. 
In Section~\ref{s-ks}, we summarize the calculations of the reflection spectrum.
In Section~\ref{s-ana}, we detail the data reduction and analysis of the observational data used in this work. 
We discuss our results in Section~\ref{s-dc}.

%%%%%%%%%%%%%%%%%%%%%%%%%%%%%%%%%%%%%%%%%%%%%%%%%%

\section{Einstein-Maxwell dilaton-axion gravity}\label{s-emda}

The Einstein-Maxwell dilation-axion (EMDA) gravity \citep{1992PhRvL..69.1006S, 2002CQGra..19.5063R, 2021MNRAS.500..481B} can be obtained by compactifying the ten dimensional heterotic string theory on a six dimensional torus $T^6$. The couplings of the metric $g_{\mu\nu}$, the $U(1)$ gauge field $A_{\mu}$, the dilaton field $\chi$, and the anti-symmetric tensor field $\mathcal{H}_{\mu\nu\alpha}$ of the third rank form the action $\mathcal{S}$ in EMDA gravity as
	\be\label{eq:action}
	\mathcal{S}&=&\frac{1}{16\pi}\int\sqrt{-g}d^4x\Bigg[ \mathcal{R}-2\partial_\nu\chi\partial^\nu\chi-\frac{1}{3}\mathcal{H}_{\rho\sigma\delta}\mathcal{H}^{\rho\sigma\delta} \nonumber\\ &&
	\qquad\qquad\qquad\quad + e^{-2\chi} \mathcal{F}_{\rho\sigma}\mathcal{F}^{\rho\sigma} \Bigg],
	\ee
	where $g$ denotes the determinant, $\mathcal{R}$ is the Ricci scalar with respect to $g_{\mu\nu}$, and $\mathcal{F}_{\mu\nu}$ is the antisymmetric Maxwell field strength tensor of the second rank, which is defined as $\mathcal{F}_{\mu\nu}=\nabla_\mu A_\nu - \nabla_\nu A_\mu$. $\mathcal{H}_{\mu\nu\alpha}$ in Eq.~\ref{eq:action} can be written as 
	\be
	\mathcal{H}_{\mu\nu\alpha} &=& \nabla_{\rho}B_{\sigma\delta}+\nabla_{\sigma}B_{\delta\rho}+\nabla_{\delta}B_{\rho\sigma} \nonumber\\ && - (A_{\rho}B_{\sigma\delta}+A_{\sigma}B_{\delta\rho}+A_{\delta}B_{\rho\sigma}),
	\ee
	where $B_ {\mu\nu}$ represents the Kalb-Ramond field, which is an antisymmetric tensor gauge field of the second rank, and when it is cyclically permuted with $A_{\mu}$, the Chern-Simons term is obtained.
	In four dimensions,  $\mathcal{H}_{\mu\nu\alpha}$ is connected to the pseudo-scalar axion field $\xi$ as
	\be
	\mathcal{H}_{\mu\nu\alpha} = \frac{1}{2}e^{4\chi}\epsilon_{\rho\sigma\delta\gamma}\partial^{\gamma}\xi.
	\ee
	We can express Eq.~\ref{eq:action} in terms of the axion field as
	\be
	\mathcal{S}&=&\frac{1}{16\pi}\int\sqrt{-g}d^4x\Bigg[ \mathcal{R}-2\partial_\nu\chi\partial^\nu\chi-\frac{1}{2}e^{4\chi}\partial_\nu\xi\partial^\nu\xi \nonumber\\ && 
	\qquad\qquad\qquad\quad + e^{-2\chi}\mathcal{F}_{\rho\sigma}\mathcal{F}^{\rho\sigma} + \xi\mathcal{F}_{\rho\sigma}\mathcal{\tilde{F}}^{\rho\sigma} \Bigg].
	\ee
	The action is varied with respect to the axion, dilaton, and Maxwell fields to obtain the corresponding equations of motion. For the dilaton field, the equation of motion can be written as
	\be
	\nabla_\mu\nabla^\mu\chi-\frac{1}{2}e^{4\chi}\nabla\mu\xi\nabla^\mu\xi+\frac{1}{2}e^{-2\chi}\mathcal{F}^2=0,
	\ee
	while that corresponding to the axion field is given by
	\be
	\nabla_\mu\nabla^\mu\xi + 4\nabla\nu\xi\nabla^\nu\xi - e^{-4\chi}\mathcal{F}_{\rho\sigma}\mathcal{\tilde{F}}^{\rho\sigma}=0.
	\ee
	When the Maxwell equations are coupled to the dilaton and axion fields, we obtain the following
	\be
	\nabla_\mu \left(e^{-2\chi\mathcal{F}^{\mu\nu}} + \xi\mathcal{\tilde{F}}^{\mu\nu} \right) &=& 0, \label{eq:maxwell1}
	\\
	\nabla_\mu \left(\mathcal{\tilde{F}}^{\mu\nu} \right) &=& 0. \label{eq:maxwell2}
	\ee
	After solving the axion, dilaton and the U(1) gauge fields, we obtain the following \citep{1992PhRvL..69.1006S, 2002CQGra..19.5063R, 2014arXiv1401.6826G}
	\be
	\xi &=& \frac{q^2}{\mathcal{M}} \frac{a\cos\theta}{r^2+a^2\cos^2\theta}, \label{eq:axion}
	\\
	e^{2\chi} &=& \frac{ r^2 + a^2\cos^2\theta }{r(r+r_2) + a^2 \cos^2\theta}, \label{eq:dilaton}
	\\
	A &=& \frac{qr}{\tilde{\Sigma}} \left( -dt + a\sin^2\theta d\phi, \right). \label{eq:gauge}
	\ee
	We can also find the non-zero components of $\mathcal{H}_{\mu\nu\alpha}$ from above equations \citep{2014arXiv1401.6826G}.
	To obtain the gravitational field equations, one needs to vary the action with respect to $g_{\mu\nu}$, yielding the Einstein's equations as
	\be
	\mathcal{G}_{\mu\nu}=\mathcal{T}(\mathcal{F}, \chi, \xi),
	\ee
	where $\mathcal{G}_{\mu\nu}$ denotes the Einstein tensor and $\mathcal{T}_{\mu\nu}$ is the energy-momentum tensor that is given by 
	\be
	\mathcal{T}(\mathcal{F},\chi,\xi) = e^{2\chi}(4\mathcal{F}_{\mu\rho}\mathcal{F}_\nu^\rho-g_{\mu\nu}\mathcal{F}^2)-g_{\mu\nu}( 2\partial_\gamma\chi\partial^\gamma\chi+\\ \frac{1}{2}e^{4\chi}\partial_\gamma\xi\partial^\gamma\xi ) + \partial_\mu\chi\partial_\nu\chi+e^{4\chi}\partial_\mu\xi\partial_\nu\xi .
	\ee

	The Kerr-Sen metric is the stationary and axisymmetric solution of the Einstein equations which takes the following form in Boyer-Lindquist coordinates \citep{1995PhRvL..74.1276G, 2013CQGra..30m5005G, 2016PhRvD..94h5007B}
	\be\label{eq:metric}
	ds^2 &=& -\left( 1-\frac{2\mathcal{M}r}{\tilde{\Sigma}} \right)dt^2+\frac{\tilde{\Sigma}}{\Delta}(dr^2+\Delta d\theta^2) \nonumber\\ && - \frac{4a\mathcal{M}r}{\tilde{\Sigma}}\sin^2\theta dtd\phi
	\nonumber\\ && + \sin^2\theta d\phi^2\left[ r(r+r_2) + a^2 + \frac{2\mathcal{M}ra^2\sin^2\theta}{\tilde{\Sigma}} \right],
	\ee
	where
	\be
	\tilde{\Sigma} &=& r(r+r_2)+a^2 \cos^2 \theta, \label{eq:sigma}
	\\
	\Delta &=& r(r+r_2)-2\mathcal{M}r+a^2. \label{eq:delta}
	\ee
	In the above equations, $\mathcal{M}$ represents the mass, $a$ is the angular momentum of the BH, and $r_2=\frac{q^2}{\mathcal{M}}e^{2\chi_0}$ is the dilaton parameter.
	The latter contains the asymptotic value of the dilatonic field $\chi_0$ and the BH electric charge $q$. The origin of this charge is not from the falling charged particles, but from the axion-photon coupling, since in the absence of an electric charge, both the axion and dilaton field strengths vanish (Eqs.~\ref{eq:axion}-\ref{eq:dilaton}). In this scenario, we get the Kerr metric from Eq.~\ref{eq:metric}. Further, it should be noted that in Eq.~\ref{eq:metric}, the BH spin comes from the axion field, since the axion field strength disappeares for a non-rotating BH (Eq.~\ref{eq:axion}). In the case of a vanishing rotation parameter, Eq.~\ref{eq:metric} reduces to a spherically symmetric pure dilaton BH, which is characterized by its mass, electric charge, and the dilaton field's asymptotic value \citep{1991PhRvD..43.3140G, 1999IJMPD...8..635Y}.

	Solving the equation $\Delta=0$, we can get the radius of the event horizon $r_H$ of Kerr-Sen space-time as
	\be
	r_H=\mathcal{M}-\frac{r_2}{2}+\sqrt{\left(\mathcal{M} - \frac{r_2}{2}\right)^2-a^2}.
	\ee

	As can be seen from Eq.~\ref{eq:sigma} and from the direct dependence of $r_2$ on the square of the electric charge, the real and positive event horizons, and hence the BH solutions, are obtained for the range $0\leq \frac{r_2}{\mathcal{M}}\leq 2$. As in \citet{2021MNRAS.500..481B}, we are interested in this range. In what follows, we will use units in which $\mathcal{M} = 1$.

%%%%%%%%%%%%%%%%%%%%%%%%%%%%%%%%%%%%%%%%%%%%%%%%%%

\section{Reflection spectra of Kerr-Sen black holes}\label{s-ks}

	X-ray reflection spectroscopy refers to the study of the reflection spectrum arising from the geometrically thin and optically thick accretion disk of a compact object \citep{2006ApJ...652.1028B, 2013mams.book.....B, 2014SSRv..183..277R, 2013MNRAS.428.2901W}. Parts of the accretion disk surface emit a spectrum similar to that of a black-body, so the entire disk has a multi-temperature black-body spectrum. Some of the thermal photons arising from the disk interact with a hot, compact, optically thin cloud of free electrons, commonly referred to as the corona. The corona geometry is still undetermined with some proposed geometries, such as the lamppost geometry, an atmosphere that surrounds the accretion disk, or an accretion flow in the plunging region between the compact object and the inner boundary of the disk. Due to interactions in the corona, photons are up-scattered through inverse Compton scattering, resulting in a power-law spectrum with a characteristic high-energy cutoff. These photons can then irradiate the accretion disk and create a reflection spectrum \citep[see, e.g.,][]{2018AnP...53000430B}.

	In the rest-frame of the emitting gas, there are rich fluorescent emission lines in the soft X-ray range ($<$ 7 keV) of the reflection spectrum, as well as the so-called Compton hump, which peaks at 20-30 keV. The iron K$\alpha$ complex around 6.4 keV for neutral or weakly ionized iron and slightly up-shifted for H-like ions is often a prominent feature of the reflection spectrum. The effects of Doppler boosting and gravitational redshift, that are different in various parts of the accretion disk, make the narrow fluorescent emission lines of the reflection spectrum broad for an observer far away. X-ray reflection spectroscopy can be a powerful technique for studying the spacetime metric around a compact object when it is complemented by a correct astrophysical model as well as high-quality data \citep{2009GReGr..41.1795S, 2013ApJ...773...57J, 2013PhRvD..87b3007B, 2019arXiv190508012A}.

	The reflection spectrum of the accretion disk that a distant observer sees can be written as
	\be
	F_o(\nu_o)&=& \int I_o(\nu_o, X, Y)d\Omega
	\nonumber\\
	&=& \int g^3 I_e(\nu_e, r_e, \theta_e)d\Omega,
	\ee
	where $I_e$ is the radiation's specific intensity measured in the rest-frame of the gas, $I_o$ is the same but measured by a distant observer, $X$ and $Y$ denote the Cartesian coordinates of image of the disk in the distant observer's plane, $d\Omega=dXdY/D^2$ is the solid angle's element that the image of the disk subtends in the observer's sky, $D$ is the distance from the observer to the source, and $r_e$ and $\theta_e$ are the emission radius and emission angle in the disk, respectively. Liouville's theorem states that $I_o=g^3 I_e$, where $g= \nu_0/\nu_e$ is the redshift factor, $\nu_e$ and $\nu_o$ are, respectively, the photon frequencies measured in the rest-frame of the emitting gas and by the distant observer.
	The transfer function formalism leads to rewriting the observed flux as~\citep{1975ApJ...202..788C}
	\be
	F_o(\nu_o) &=&\frac{1}{D^2}\int_{r_{in}}^{r_{out}}\int_{0}^{1} \pi r_e \frac{g^2}{\sqrt{g^*(1-g^*)}}
	\nonumber\\
	&& \qquad \times f(\nu_e, r_e, \theta_e) \ I_e(\nu_e, r_e, \theta_e) \ dg^* \ dr_e,
	\ee
	where $r_{in}$ ($r_{out}$) represents the inner (outer) edge of the accretion disk, and $f$ is the transfer function defined as
	\be
	f(\nu_e, r_e, \theta_e) = \frac{g\sqrt{g^*(1-g^*)}}{\pi r_e} \left| \frac{\partial(X, Y)}{\partial (g^*, r_e)} \right|,
	\ee
	where $| \partial(X, Y) / \partial (g^*, r_e) |$ denotes the Jacobian of the coordinate transformation from the Cartesian coordinates of the distant observer to the disk coordinates ($g^*, r_e$), $g^*=g^*(r_e, \theta_e)$ is the relative redshift factor. It is defined as 
	\be
	g^*=\frac{g-g_{min}}{g_{max} - g_{min}} \in [0,1].
	\ee
	where $g_{max}=g(r_e, \iota)$ ($g_{min}=g(r_e, \iota)$) is the maximum (minimum) value of the redshift factor $g$ for a family of photons originating from the radial coordinate $r_e$ and detected at a far distance with the viewing angle $\iota$.

	The transfer function is determined by the spacetime metric and the distant observer's viewing angle, and includes relativistic effects such as Doppler boosting, gravitational redshift, and light bending. For more details, refer to \citet{2017ApJ...842...76B, 2019ApJ...878...91A, 2020ApJ...899...80A}. In this work, a general relativistic ray-tracing code is used to calculate the Jacobian $| \partial(X, Y) / \partial (g^*, r_e) |$ and then the transfer function for the Kerr-Sen metric (Eq.~\ref{eq:metric}).

\begin{figure}
\begin{center}
\includegraphics[width=0.49\textwidth,trim={0cm 0cm 0cm 0cm},clip]{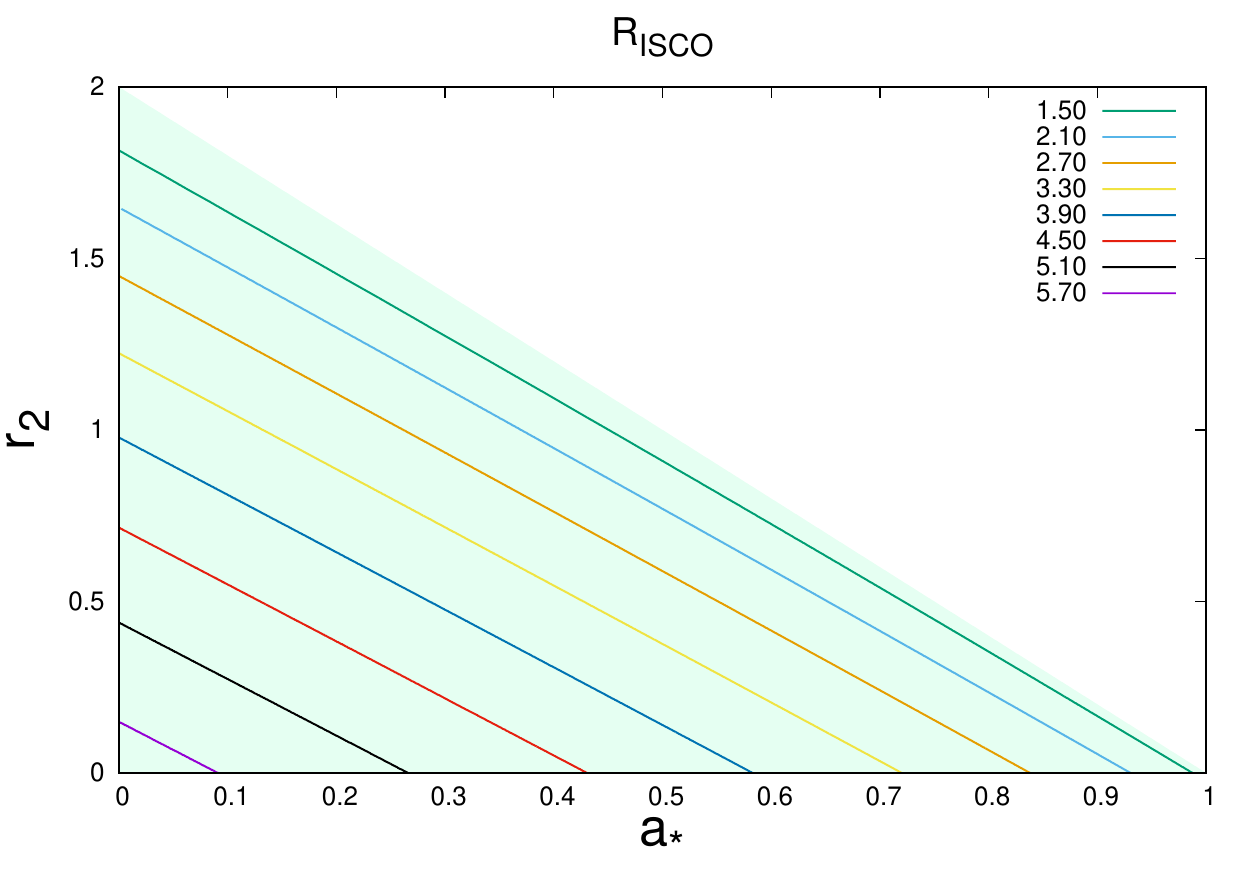}
\end{center}
\vspace{-0.2cm}
\caption{Contour levels of the radial coordinate of the ISCO radius in the plane $a_*$ vs $r_2$. $R_{ISCO}$ in units of $\mathcal{M}=1$. \label{f-isco}}
%\vspace{0.4cm}
\end{figure}

\begin{figure}
\begin{center}
\includegraphics[width=0.49\textwidth,trim={0cm 0cm 0cm 0cm},clip]{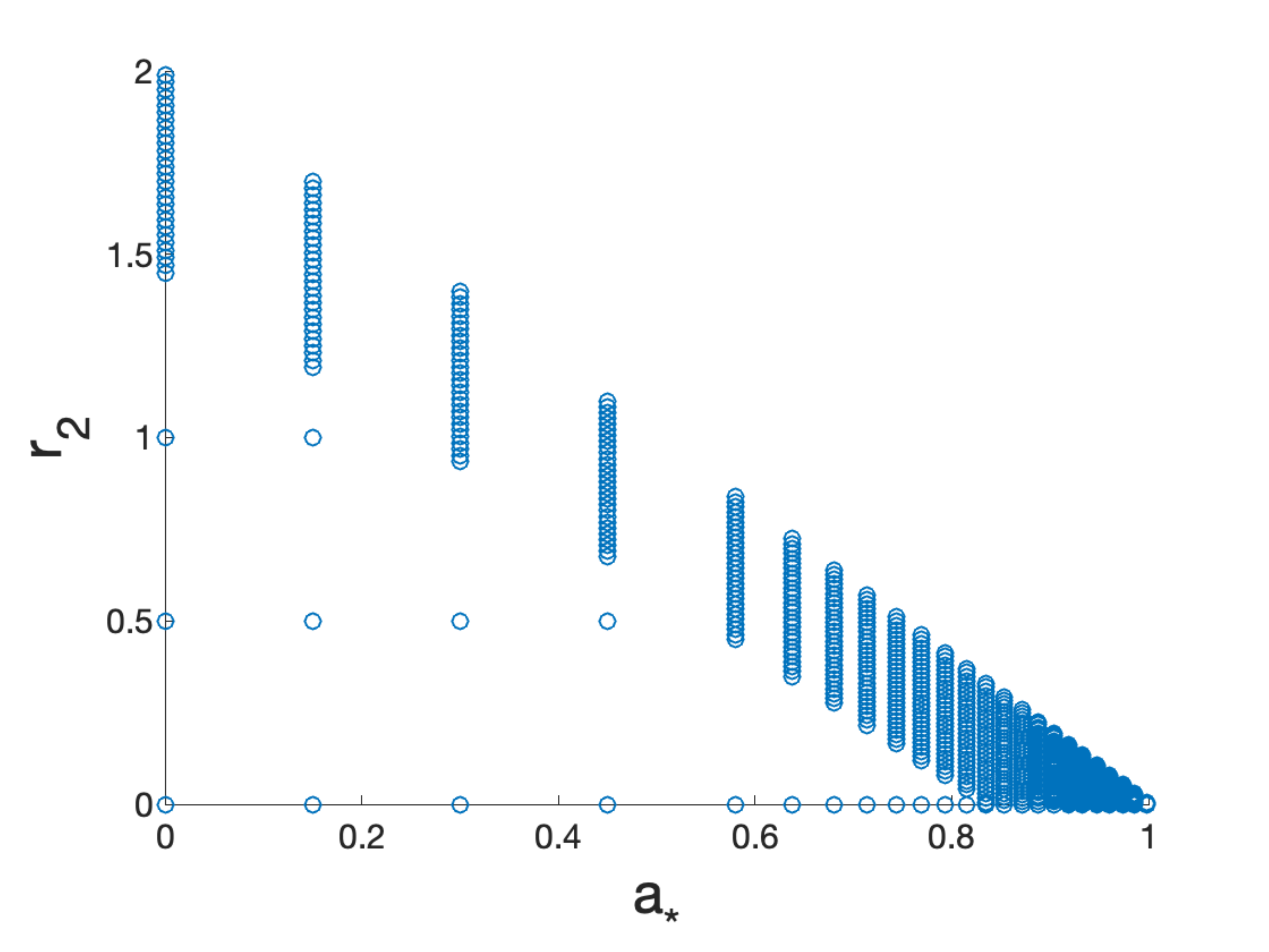}
\end{center}
\vspace{-0.2cm}
\caption{Grid points of the FITS file of the transfer function on the plane spin parameter $a_*$ vs deformation parameter $r_2$. \label{f-grid}}
%\vspace{0.4cm}
\end{figure}

	The sequence of calculations is as follows. First, we discretize the accretion disk into 100 emitting rings, which range from $R_{ISCO}$ to $1000$, where $R_{ISCO}$ is the radius of the innermost stable circular orbit (ISCO). Fig.~\ref{f-isco} shows the values of $R_{ISCO}$ depending on $a_*$ and $r_2$, where $a_*=J/\mathcal{M}^2$ is the dimensionless spin parameter of the Kerr-Sen BH. In each emitting ring, we mark points on the accretion disk that correspond to 40 equidistant values of $g^*$, and calculate the transfer functions for each of these points. This calculation is repeated for each set of ($a_*, r_2, \iota$). The grid of viewing angle $\iota$ is selected as in \citet{2017ApJ...842...76B, 2019ApJ...878...91A, 2020ApJ...899...80A}, while the grid points on the plane ($a_*, r_2$) are illustrated in Fig.~\ref{f-grid}. The values of the transfer function are tabulated in the FITS file, and then used in the \textsc{relxill\_nk} model for arbitrary geometry to calculate the single line profile and full reflection spectra \citep{2017ApJ...842...76B, 2019ApJ...878...91A, 2020ApJ...899...80A}.

%%%%%%%%%%%%%%%%%%%%%%%%%%%%%%%%%%%%%%%%%%%%%%%%%%

\section{Observational constraints}\label{s-ana}

We will now employ the model developed using the EMDA metric for analyzing X-ray 
data from an X-ray binary to measure the parameter $r_2$. In this section, 
we will discuss the source, observation used, and the analysis. 

The observation analyzed in this work is the 2019 \textsl{NuSTAR} observation of the X-ray binary EXO~1846--031 (ObsID 90501334002). 
First observed by \textsl{EXOSAT} in 1985 \citep{1985IAUC.4051....1P}, EXO 1846--031 was later confirmed 
as a low mass  X-ray binary \citep{1993A&A...279..179P}. After about 9 years, \cite{1994IAUC.6096....1Z} reported 
the next outburst of this source observed with \textsl{CGRO}/BATSE. This source had its third outburst after
about 25 years detected by \textsl{MAXI} recently in July 2019 \citep{2019ATel12968....1N}. After the 2019 detection, several
telescopes have observed this source. The focus of this work is a \textsl{NuSTAR} observation, first analyzed by \citet{2020ApJ...900...78D}.
This observation has very simple spectra without any warm or neutral absorption from winds and displays strong
reflection features. This particular observation has also been analyzed for different spacetime metrics 
\citep{2021arXiv210204695T, 2020arXiv201210669T, 2021arXiv210110100A}. In all cases, the Kerr hypothesis
is within the confidence contours which indicates the BH to be described by the Kerr solution. {\sc relxill\_nk} provides a suitable description 
of the relativistic spectra and the emissivity from the accretion disk with a broken power-law.  
The inner emissivity is very high ($>7$) and the outer emissivity is very low ($\sim 1$) which suggests 
the corona to be compact and close to the BH. The BH in EXO~1846--031 
has an iron line that is asymmetrically broadened implying a high spin parameter and that
the radiation is primarily coming from the inner radii of the accretion disk. The angle between the observer's line of sight and the accretion disk's angular momentum, i.e.~the inclination angle, is also seen to be high, which along with other features, make the relativistic
effects more prominent. \textsl{NuSTAR}, having impressive sensitivity  and ability to observe bright sources without pile-up, is the 
best observatory today for observing disk reflection features of bright sources throughout the entire energy range of 3.0-79.0 keV,
covering the iron line and the Compton hump features. All these factors justify the use of this particular
observation of this source. 

\subsection*{Data Reduction}
EXO~1846--031 was observed for about 22 ks by both \textsl{NuSTAR} \citep{2013ApJ...770..103H} detectors; Focal plane module (FPM) A and B  on 
29th of August in the year of 2019. The raw data obtained from HEASARC website is processed into cleaned event files
using calibration database (CALDB) v20200912 and NUPIPELINE routine of NUSTARDAS analysis software which is
distributed as a part of high energy analysis software HEASOFT package. The source region of 180 arcsec is selected around the source from 
the cleaned event file. A background region of the same size as that of the source is taken from the region as far as
possible from the source at the same detector. Source spectra, background spectra, response and ancillary files are
generated using the NUPRODUCT routine of NUSTARDAS. For the chi-squared statistics to be applicable, the source spectra
is binned to have a minimum of 30 counts per bin.

\subsection*{Spectral Analysis}
X-ray spectral analysis package XSPEC v12.11.1 is used for the analysis done in the paper.
We used abund WILMS \citep{2000ApJ...542..914W} and cross-section VERNER \citep{1996ApJ...465..487V}
for all the calculations. The energy range of 3.0-79.0 keV is analyzed for both FPMA and FPMB 
detectors.

Since this observation is relatively simple without any absorption and shows strong reflection
features, we start the fit with the model {\sc tbabs*relxill\_nk}. {\sc tbabs} describes the galactic absorption
and has only one parameter; column density ($n_H$) along the line of sight. $n_H$ is set to be free while
fitting the spectra. {\sc relxill\_nk} describes the power-law and reflection components. We assume that the inner
edge of the accretion disk is at the ISCO and the outer radius is set to 
$400~r_g$ where $r_g$ is the gravitational radius. The emissivity profile is modeled with a broken power-law
emissivity profile with three parameters; inner emissivity ($q_{in}$), outer emissivity ($q_{out}$) and break
radius ($R_{br}$) where the emissivity changes. The reflection fraction is frozen to 1 because this parameter in
{\sc relxill\_nk}, unlike other reflection models where it has geometrical meaning, is a scale factor which is found 
to be degenerate with its normalization. The other parameters of {\sc relxill\_nk} are the spin parameter of the BH $a_*$
(its value lies between -1 and 1), the iron abundance (relative to its solar abundance), ionization of the 
disk ($\xi$), and the inclination angle of the disk with respect to line of sight of observer. The power-law components are 
described by two parameters; power-law of the flux from the corona which illuminates the accretion disk ($\Gamma$) 
and the cutoff energy of the power-law ($E_{cut}$). We also try various flavors of {\sc relxill\_nk} to fit this data. Based
on the chi-squared statistic, we choose the standard {\sc relxill\_nk} for further analysis.

\begin{figure}
\begin{center}
\includegraphics[width=0.45\textwidth,trim={0cm 0cm 0cm 0cm},clip]{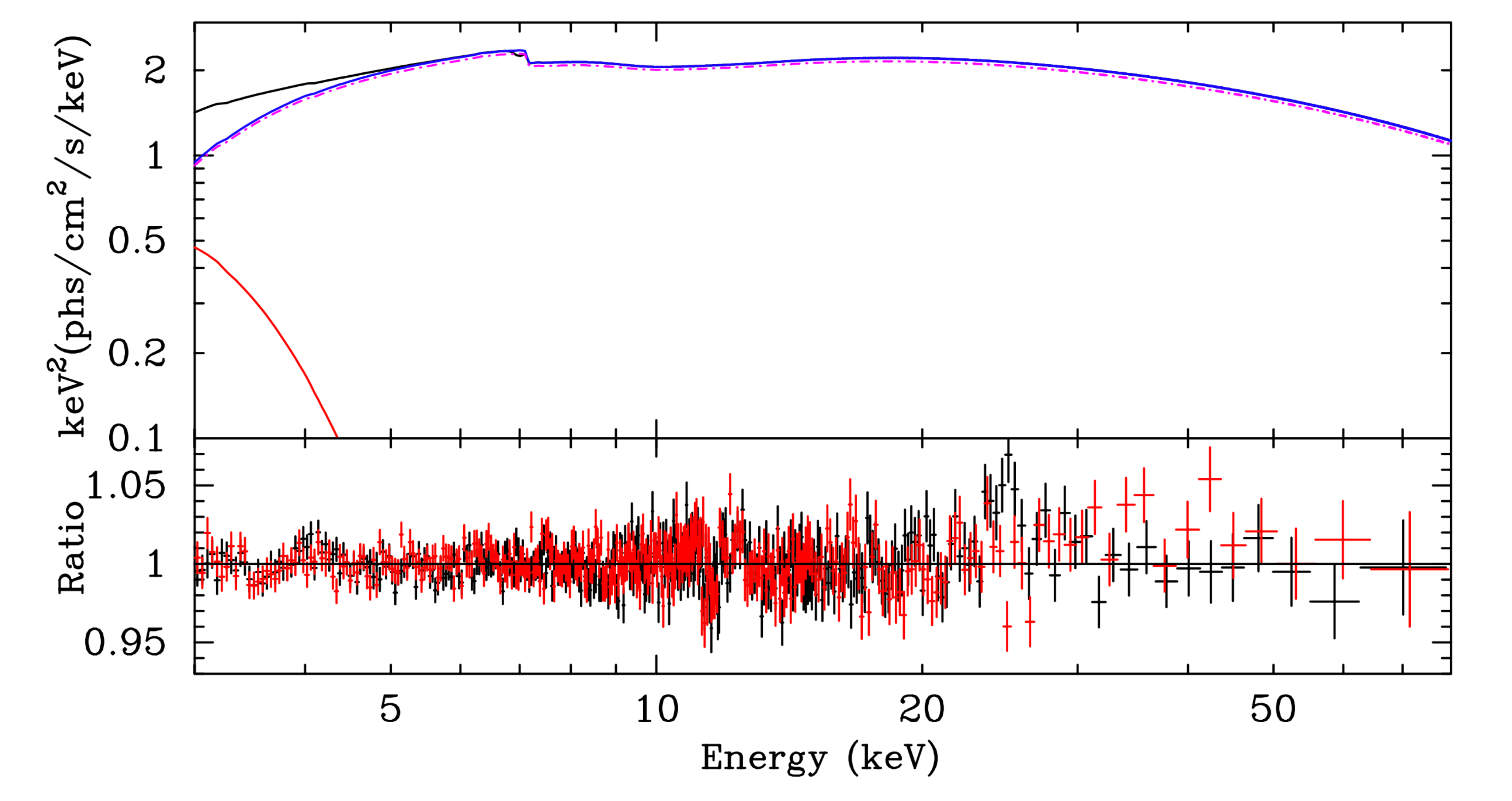}
\end{center}
\vspace{-0.2cm}
\caption{Best fit model (upper panel) and data to the best-fit model ratio for EMDA metric for \textsl{NuSTAR} observation of EXO~1846--031. In the upper panel, the red color corresponds to {\sc diskbb}, the blue color corresponds to {\sc relxill\_nk}, and the black curve denotes the total model {\sc tbabs*(diskbb+relxill\_nk+gaussian)}. The dashed-dotted magenta curve corresponds to {\sc relxill\_nk} of the best-fit model with $r_2$ set at the maximum value allowed by the spin parameter. In the lower panel, the black color denotes the ratio of FPMA and the red color denotes that of FPMB.\label{f-ratio}}
\end{figure}

The residuals at low energies is fixed by adding another additional model {\sc diskbb} which models the continuum component
from the infinitesimally thin accretion disk. Some absorption residuals can also be
seen around 7 keV which is modeled with a Gaussian line. This feature can be explained 
by absorption by winds around the disk at a relatively high inclination angle. 
\cite{2014ApJ...784L...2K} reported similar features 
for \textsl{NuSTAR} observation of another X-ray binary 4U~1630--472. The top and bottom panel of Fig.~\ref{f-ratio} represents the best fit model 
and the ratio of the data relative to the best fit model, respectively. 
We note that we nicely recover the results of \citet{2020ApJ...900...78D}, including the absorption line at 7~keV.

At lower energies, the residuals for the detectors are very different. It is most probably due to an instrumental
issue mentioned in \citet{2020arXiv200500569M}. We follow the procedure given in \citet{2020ApJ...900...78D}
and \citet{2021arXiv210204695T}. We first fit the spectra ignoring 3.0-7.0 keV and 
fix the cross-calibration constant. Then, we noticed the 3.0-7.0 keV energy range and added the multiplicative
table (mtable) described in \citet{2020arXiv200500569M} which has only one parameter $cov_{frac}$. We fix the $cov_{frac}$
of FPMB  to 1 and vary the $cov_{frac}$ of FPMA. So, the best fit model used in this work is 
{\sc constant*mtable*tbabs*(diskbb+relxill\_nk+gaussian)}

\begin{figure*}
\begin{center}
\includegraphics[width=0.98\textwidth,trim={0cm 0cm 0cm 0cm},clip]{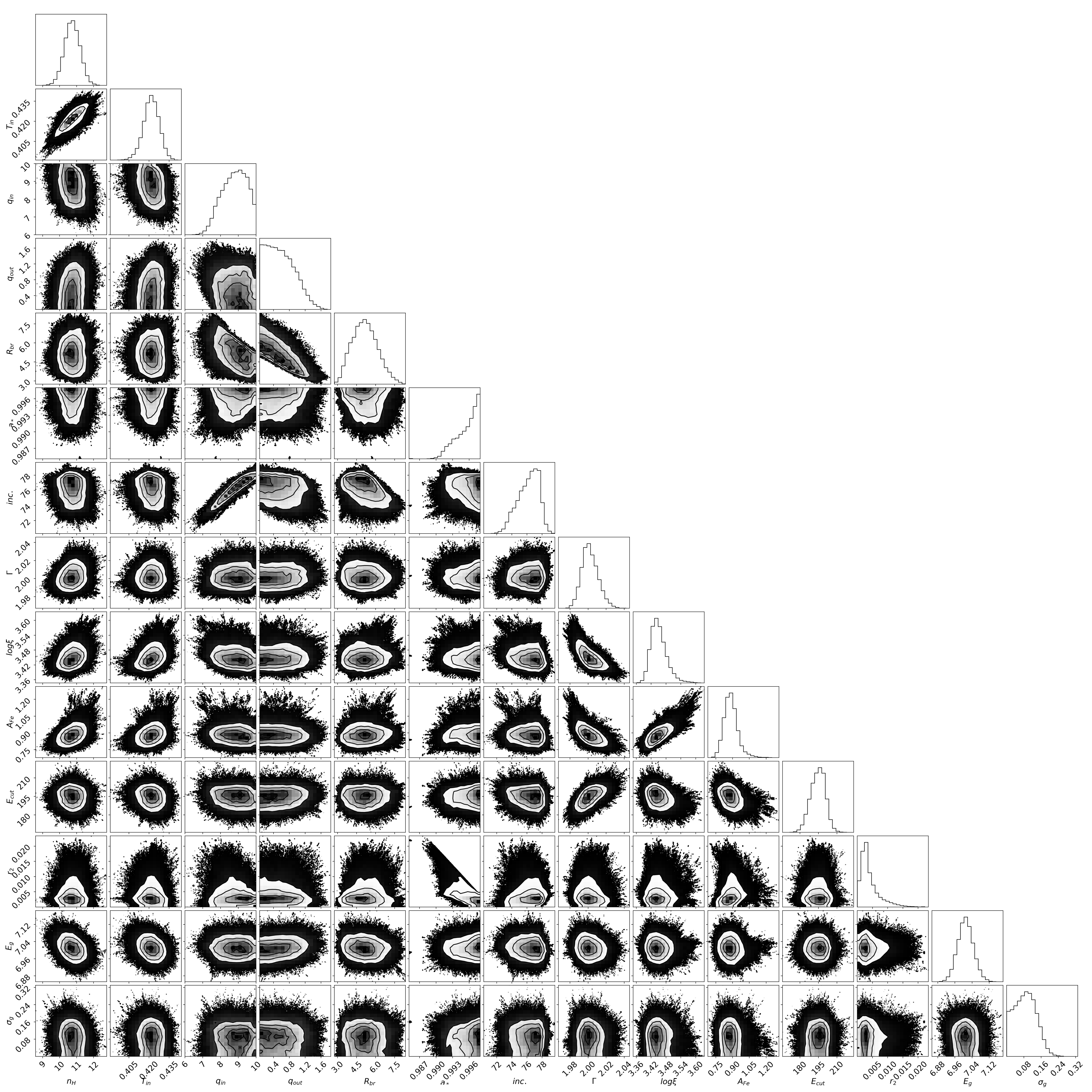}
\end{center}
\vspace{-0.2cm}
\caption{The corner plot for the parameters of the best-fit model for the Observation used in this work (excluding constants and normalization of models). The 2D plots shows 1-, 2-, and 3-$\sigma$ confidence contours. \label{f-mcmc}}
%\vspace{0.4cm}
\end{figure*}

\begin{figure}
\begin{center}
\includegraphics[width=0.49\textwidth,trim={0cm 0cm 0cm 0cm},clip]{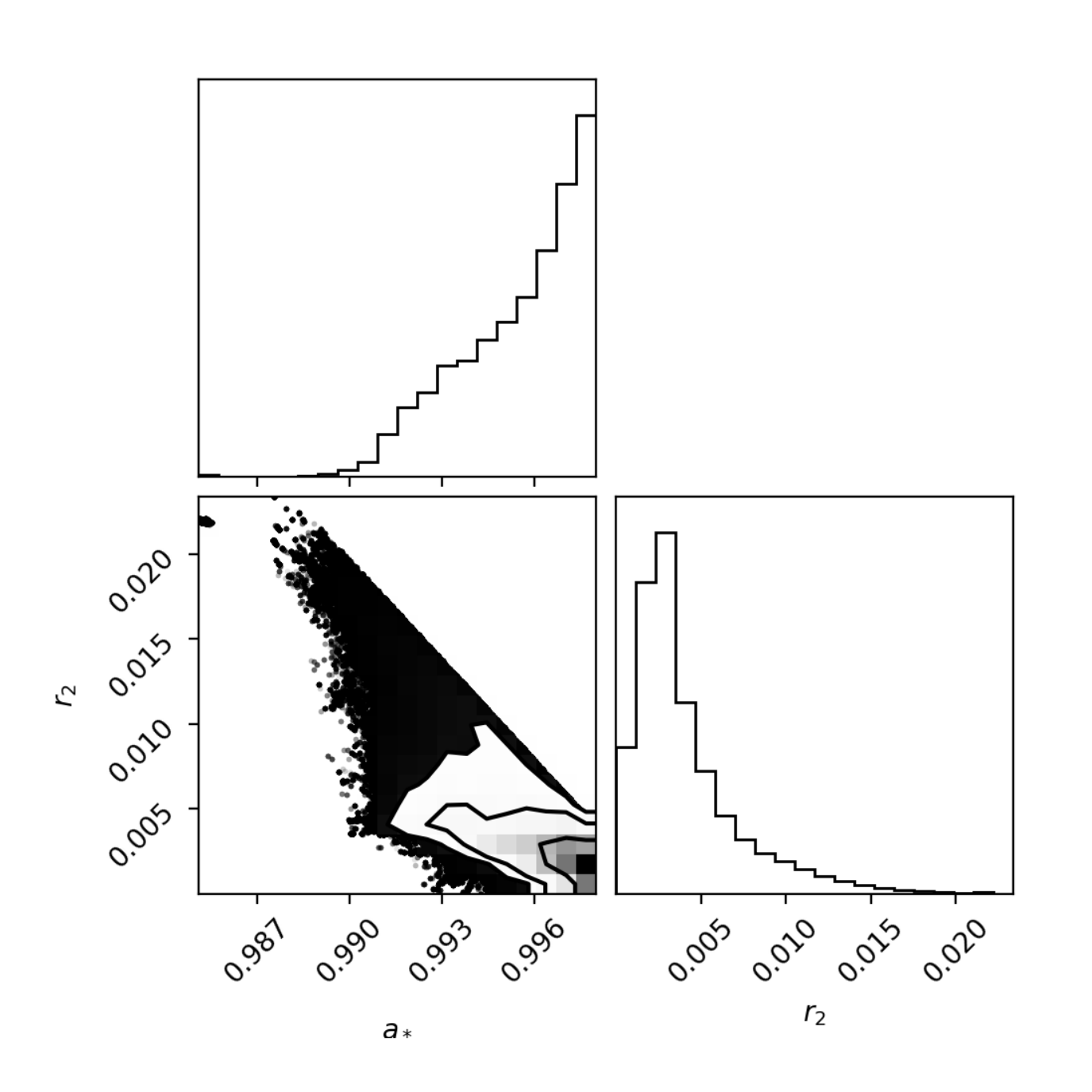}
\end{center}
\vspace{-0.5cm}
\caption{The histograms for spin and deformation parameter space for the best fit model. The 2D plot represents the 1-, 2-, and 3-$\sigma$ confidence contours.  \label{f-mcmczoom}}
%\vspace{0.4cm}
\end{figure}

We employed chi statistics to determine the best fit parameters which serves as a prior distribution for Markov chain
Monte Carlo (MCMC) analysis. We used the python script by Jeremy Sanders \footnote{the script is available at \href{https://github.com/jeremysanders/xspec_emcee}{https://github.com/jeremysanders/xspec\_emcee}.}
which uses an XSPEC model file to run the chains
by employing emcee, which is a MCMC sampler employing the Goodman \& Weare algorithm. The chains were run for 300 walkers 
with 20000 iterations and initial 2000 burn-in steps. Thus, there are a total of $6\times 10^6$ samples. The values of
the parameters from the MCMC represents the median across the samples. The error is calculated for 90 percent confidence 
intervals across the whole parameter chain. Please note that these errors are only statistical. The 90\% error on the parameters for the best fit model obtained from MCMC simulations are given in Table~\ref{table}. 
The corner plot resulting from MCMC analysis is shown in Fig.~\ref{f-mcmc}. The zoomed corner plot between spin and $r_2$ is shown 
in Fig.~\ref{f-mcmczoom}.

\begin{table} 
\centering
\renewcommand\arraystretch{1.3}{
\begin{tabular}{lc}\hline\hline
Parameter &EXO~1846--031 \\ \hline \hline
$cov_{frac}$&$0.881^{+0.015}_{-0.015}$\\
$n_{H}$ [$10^{22}$~cm$^{-2}$] & $10.7^{+0.7}_{-0.7}$\\
$q_{in}$&    $8.8^{+1.2}_{-1.0}$\\
$q_{out}$&   $0.6^{+0.6}_{-0.4}$\\
$R_{br}$ [$r_g$]&   $5.1^{+1.8}_{-1.5}$\\
$a_*$& $0.996^{+0.001}_{-0.004}$\\
$\iota$ [deg] &$76.3^{+1.7}_{-2.7}$\\
$\Gamma$ & $2.001^{+0.017}_{-0.013}$\\
$\log\xi$ [erg~cm~s$^{-1}$] & $3.45^{+0.18}_{-0.03}$\\
$A_{Fe}$ & $0.88^{+0.11}_{-0.09}$\\
$r_2$& $0.003^{+0.008}_{}$\\
$E_{cut}$ [keV]&$195^{+9}_{-10}$\\
$N_{relxill\_nk}$ [$10^{-3}$] & $9.5^{+0.5}_{-0.4}$\\
$T_{in}$ [keV] &$0.422^{+0.009}_{-0.010}$\\
$N_{diskbb}$ [$10^{4}$] &$4.1^{+0.6}_{-0.4}$\\
$E_{g}$&$7.01^{+0.07}_{-0.06}$\\
$\sigma_{g}$&$0.09^{+0.08}_{-0.08}$\\
$N_{g}$ [$10^{-4}$] &$-5.1^{+1.4}_{-1.7}$\\ 
\hline
$\chi^2$/dof & 2748.50/2594 = 1.0596 \\
\hline\hline
\end{tabular}}
\vspace{0.3cm}
 \caption{Best fit parameters and the corresponding 90\% error for the best fit model for the 2019 \textsl{NuSTAR} observation of EXO~1846--031. See text for more details.} \label{table}
\label{tab1}
\end{table}

%%%%%%%%%%%%%%%%%%%%%%%%%%%%%%%%%%%%%%%%%%%%%%%%%%

\section{Discussion and conclusions}\label{s-dc}

EMDA gravity is a string-inspired model arising from the low energy effective action of heterotic string theory. In this model, the spacetime around rotating BHs is described by the Kerr-Sen metric, where we have a BH dilaton charge and a BH angular momentum induced by the axion field. These BHs are different from the Kerr BHs of GR, and this fact opens the possibility of testing EMDA gravity with astrophysical observations of BHs, as already explored by other authors \citep{2007PhRvD..75b3006G, 2018PhRvD..97b4003A, 2016PhRvD..94h4025Y, 2008PhRvD..78d4007H, 2020arXiv200212786N,2021MNRAS.500..481B}. In particular, \citet{2021MNRAS.500..481B} derived the constraints $r_2 < 0.1$ ($r_2 \ge 0$ is the BH dilaton charge and we recover the Kerr BHs of GR for $r_2 = 0$) from the analysis of the optical luminosity for 80 Palomar-Green quasars.

In this work, we have constructed a reflection model in the Kerr-Sen spacetime and we have constrained the BH dilation charge $r_2$ from the analysis of a reflection-dominated \textsl{NuSTAR} spectrum of the BH binary EXO~1846--031. Our constraint, which is the main result of this work, is (90\% CL)
\be\label{eq-f-c}
r_2 < 0.011 ,
\ee
which is an order of magnitude more stringent than the constraint inferred in \citet{2021MNRAS.500..481B}. 
Our analysis essentially recovers the results presented in \citet{2020ApJ...900...78D}, in which it was assumed the Kerr background, and therefore the Kerr metric is enough to explain the data.

While there are several simplifications in our model, we stress that our constraint on $r_2$ is robust. We analyzed the \textsl{NuSTAR} observation assuming that the inner edge of the accretion disk is at the ISCO radius, as it is common in X-ray reflection spectroscopy when we want to measure BH spins. While there is still some debate whether accretion disks in the hard states are truncated or not \citep[see, e.g.,][and references therein]{{2020arXiv201104792B}}, in our case we find a BH spin close to the maximum value allowed by the model, so the data require that the inner edge of the accretion disk is close to the BH and the truncation of the disk, if any, can only be very modest. The model employs an infinitesimally thin disk, while the disk has a finite thickness. However, \citet{2021arXiv210204695T} analyzed this \textsl{NuSTAR} observation with a model with a disk of finite thickness without finding any important bias in the estimate of the model parameters. The impact of a non-constant ionization parameter over the accretion disk was investigated in \citet{2021arXiv210110100A}, concluding that the estimate of the deformation parameter is not affected by the ionization gradient. Our reflection model ignores the returning radiation, namely the radiation emitted by the disk and returning to the disk because of the strong light bending near the BH. While the effect can indeed induce important systematic uncertainties in some measurements when the inclination angle of the accretion disk is low  \citep[see, e.g., ][]{2020arXiv200615838R}, we expect only a weak impact on the estimate of $r_2$ in our case, as the viewing angle is definitively high. Our reflection spectrum is calculated assuming a disk electron density $n_{\rm e} = 10^{15}$~cm$^{-3}$, which is probably too low for the accretion disk in an X-ray binary \citep[see, e.g, ][]{2019MNRAS.484.1972J,2019MNRAS.489.3436J}. However, if we employ a higher disk electron density, the reflection spectrum mainly changes below a few keV, which is not covered by the \textsl{NuSTAR} data, and the impact of higher values of $n_{\rm e}$ is thought to be quite weak for the estimate of the BH spin or possible deformation parameters like $r_2$ \citep{2019MNRAS.484.1972J,2019MNRAS.489.3436J}.
Lastly, we note that our analysis assumes the simplest continuum model, a power-law with an exponential cutoff, which works very well for these data. However, different continuum assumptions are possible, and this can also somewhat affect the exact constraint on $r_2$.

Our constraint in Eq.~\ref{eq-f-c} can be improved if, for example, we have the possibility of analyzing a brighter source and of simultaneously fitting a strong reflection component and a prominent thermal spectrum~\citep[see ][]{2021ApJ...907...31T}. Future X-ray missions, like \textsl{XRISM} \citep[projected launch in $\sim$2022;][]{2020SPIE11444E..22T} and \textsl{Athena} \citep[projected launch in $\sim$2034;][]{2013arXiv1306.2307N} will have microcalorimeters with superb energy resolution in the iron line region and they are expected to provide much better reflection measurements than those possible today.

%%%%%%%%%%%%%%%%%%%%%%%%%%%%%%%%%%%%%%%%%%%%%%%%%%

\section*{Acknowledgements}

This work was supported by the Innovation Program of the Shanghai Municipal Education Commission, Grant No.~2019-01-07-00-07-E00035, the National Natural Science Foundation of China (NSFC), Grant No.~11973019, and Fudan University, Grant No.~JIH1512604.

\section*{Data availability}

The \textsl{NuSTAR} raw data analyzed in this work are available to download at the HEASARC Data Archive website\footnote{\href{https://heasarc.gsfc.nasa.gov/docs/archive.html}{https://heasarc.gsfc.nasa.gov/docs/archive.html}}.

%%%%%%%%%%%%%%%%%%%%%%%%%%%%%%%%%%%%%%%%%%%%%%%%%%



%\bibliographystyle{mnras}


%%%%%%%%%%%%%%%%%%%%%%%%%%%%%%%%%%%%%%%%%%%%%%%%%%

%%%%%%%%%%%%%%%%% APPENDICES %%%%%%%%%%%%%%%%%%%%%

% Don't change these lines
\bsp	% typesetting comment
\label{lastpage}
\end{document}